\documentclass[epj,final]{svjour}

\usepackage{graphicx}
\usepackage{dcolumn}   
\usepackage{bm}        
\usepackage{amssymb}   
\usepackage{textcomp}
\usepackage{ucs}
\usepackage{hyperref}
\usepackage{amsmath}
\usepackage{color}
\usepackage{footnote}

\begin{document}

\title{Attenuation of vacuum ultraviolet light in liquid argon}

\author{A. Neumeier\inst{1} \and M. Hofmann\inst{1,4} \and L. Oberauer\inst{1} \and W. Potzel\inst{1} \and S. Sch\"onert\inst{1} \and T. Dandl\inst{2} \and T. Heindl\inst{2} \and A. Ulrich\inst{2}\thanks{\emph{Andreas Ulrich:} andreas.ulrich@ph.tum.de 
} \and \\J. Wieser\inst{3}
}                     
\institute{Technische Universit\"at M\"unchen, Physik-Department E15, James-Franck-Str. 1, D-85748 Garching, Germany \and Technische Universit\"at M\"unchen, Physik-Department E12, James-Franck-Str. 1, D-85748 Garching, Germany \and excitech GmbH, Branterei 33, 26419 Schortens \and now at: KETEK GmbH, Hofer Str. 3, 81737 M\"unchen, Germany
}

\date{Published in Eur. Phys. J. C (2012)}

\abstract{
The transmission of liquid argon has been measured, wavelength resolved, for a wavelength interval from 118 to 250\,nm. The wavelength dependent attenuation length is presented for pure argon. It is shown that no universal wavelength independent attenuation length can be assigned to liquid argon for its own fluorescence light due to the interplay between the wavelength dependent emission and absorption. A decreasing transmission is observed below 130\,nm in both chemically cleaned and distilled liquid argon and assigned to absorption by the analogue of the first argon excimer continuum. For not perfectly cleaned argon a strong influence of impurities on the transmission is observed. Two strong absorption bands at 126.5 and 141.0\,nm with approximately 2 and 4\,nm width, respectively, are assigned to traces of xenon in argon. A broad absorption region below 180\,nm is found for unpurified argon and tentatively attributed to the presence of water in the argon sample.
}           

\maketitle
\section{Introduction}
\label{sec:Introduction}
Liquid rare gases have been proven to be excellent media for particle detection with fluorescence detectors due to their high fluorescence efficiency. Liquid argon and xenon, in particular, are used for rare event search in astroparticle physics and benefit from the outstanding scintillation properties of liquid rare gases. Several already existing or future experiments use argon as detector medium, both in scintillation counters and time projection chambers, covering nearly all present fields of astroparticle physics like direct search for dark matter particles (MiniCLEAN \cite{MiniCLEAN}, DEAP-3600 \cite{DEAP/CLEAN}, ArDM \cite{ArDM_1,ArDM_2}, WARP \cite{WARP}, DarkSide \cite{DarkSide}, and DARWIN \cite{DARWIN_1,DARWIN_2}), neutrino physics (ICARUS \cite{ICARUS} and GLACIER \cite{Rubbia_CP_Violation}), or the search for neutrinoless double-beta decay (GERDA \cite{GERDA} uses a liquid argon veto). 
        
        Since some of the detector systems mentioned above use large quantities of liquid argon the scintillation light may have to traverse long distances from the position where it is produced to the detector. It is therefore important to study the attenuation of vacuum ultraviolet (VUV) light in liquid argon. However, an intensive literature search returned no publication showing a wavelength resolved attenuation length measurement for pure liquid argon. Ishida et al. \cite{Ishida_Abschwaechlaenge} measured, among other noble gases and admixtures, the wavelength integrated attenuation length of pure liquid argon for its own scintillation light. The measuring principle was based on recording the amount of scintillation light (obtained by irradiating liquid argon with a heavy ion beam) at seven different distances between 137\,mm and 737\,mm from the light detector. For that purpose, liquid argon was contained in a steel cell with seven entrance windows for the heavy ion beam. Light detection was performed at the end of the steel cell via a Pyrex$^{\circledR}$ window coated with sodium salicylate, which shifted the VUV scintillation light into visible light to enable detection with a regular photomultiplier. A wavelength integrated attenuation length of (66$\pm$3)\,cm in pure liquid argon was obtained in this experiment. 
        
        Recently we have measured the emission spectrum of liquid argon in the vacuum ultraviolet spectral range. The main result of this study \cite{Heindl_1} will be shown again here (section \ref{subsec:Pure_liquid_argon}). It is obvious that the wavelength integrated measuring principle cannot disentangle the influence of possible impurities like water vapor, oxygen, or other noble gases, which have a huge effect on the transmission. It has also to be noted that impurities, depending on their concentrations, affect both the emission spectrum as well as the transmission spectrum. Any wavelength integrated measuring principle can therefore only measure a combined effect of emission and transmission, and consequently the obtained "attenuation length" strongly depends on impurities. We will show that an attenuation length (for wavelength-integrated light) in the sense of an exponential decay of intensity and a reduction of light to $\frac{1}{e}$ cannot be applied due to the wavelength dependence of the absorption. We will also show that traces of other rare gases, xenon in particular in the case of liquid argon, can contribute significantly to light absorption. Since Ishida et al. \cite{Ishida_Abschwaechlaenge} used the same experimental setup to measure all noble gases and admixtures, the probability of a noble gas contamination is high in these experiments but cannot be identified due to the wavelength-integrated data taking method. Wavelength resolved reference data for the absorption of xenon-doped liquid argon are available from a measurement by Raz and Jortner in 1970 \cite{Jortner_1970} and will be used to estimate the xenon impurity concentration in our experiments. 
        
        Here we follow the concept that the optical properties of liquid rare gases are similar to those of dense gases. From the latter some information is available to predict the attenuation. Detailed studies of the issue have been performed in the context of VUV excimer lasers of pure rare gases \cite{Photoabsorption_Cross_Section_Krypton_Xenon}. Attenuation on the long wavelength side of the resonance lines of dense krypton and xenon gas has been described in ref.\,\cite{Photoabsorption_Cross_Section_Krypton_Xenon}  in that context. Both absorption due to optical transitions starting from the van der Waals minimum of the lower potential curve in the region of the so-called first excimer continuum and scattering due to increasing fluctuations of the index of refraction when approaching the resonance lines from longer wavelengths have been identified in that reference as the reason for light attenuation extending far out into the red wings of the resonance lines. Rayleigh scattering in rare gas liquids has also more recently been discussed in ref.\,\cite{Rayleigh_scattering_length_calculation} resulting in a scattering length of 90\,cm. Here we assume that the processes discussed in these publications are also relevant for the attenuation in liquid argon for wavelengths close to the resonance lines of argon at 104.82\,nm 106.67\,nm \cite{NIST}, and longer wavelengths.

        After a description of experimental details the VUV transmission of the best quality liquid argon achieved in this study will be presented. Then, it will be shown that traces of xenon in liquid argon lead to significant absorption bands, in part exactly at the peak of the emission band of liquid argon, and the data will be discussed using the results from ref.\,\cite{Jortner_1970}. Attenuation by impurities is a well known effect in the vacuum ultraviolet spectral region, and rare gas impurities are particularly critical because they are not removed by regular chemical gas purification techniques. It will be shown that these chemical purification steps are mandatory to remove regular impurities like oxygen and water vapour.

\section{Experimental details}

        \subsection{Experimental setup}

\begin{figure} 
            \includegraphics[width=\columnwidth]{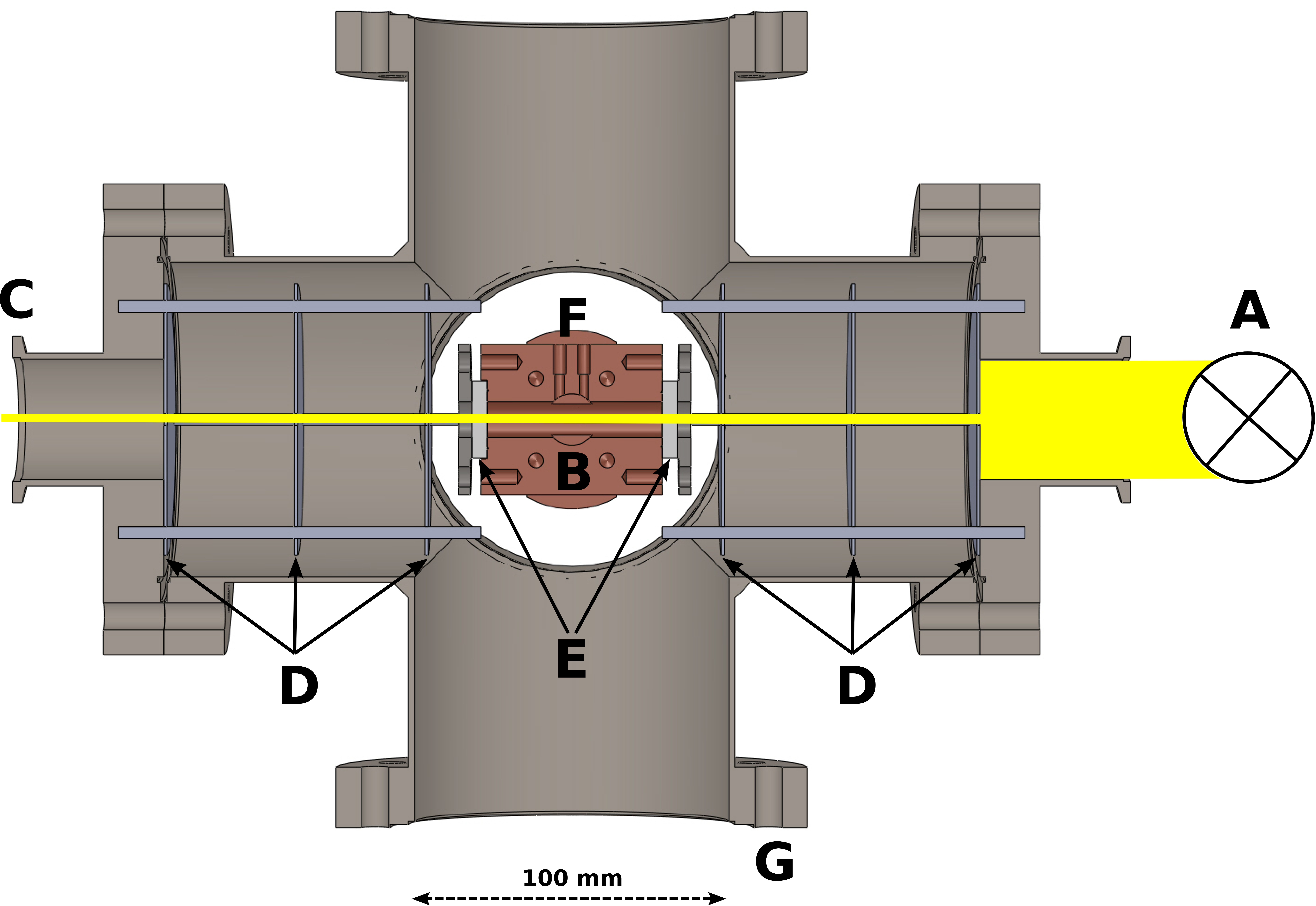} 
            \caption{\textit{Transmission measurements in liquid argon were performed by sending light from a deuterium lamp (A), which was attached to an evacuated CF-100 crosspiece (outer cell, G), through a 58\,mm long inner cell made from copper (B) containing liquid argon. Apertures (D), both in front and behind the inner cell, were used to avoid stray light. The inner cell (B) was attached to a liquid nitrogen dewar for cooling (not shown) and was sealed with MgF$_{2}$ windows (E) on the optical axis. Gas in- and outlet were provided via the connections (F). The evacuated inner cell kept at the same temperature was used for recording reference spectra. A f=30\,cm VUV spectrometer was placed at the light output (C) and used to record the spectra.}} 
            \label{fig:Strahlengang} 
        \end{figure}

        The concept of the experiment was to use a very basic optical setup to avoid complications in the interpretation of the results for example due to changes in the geometry when the gas cell is cooled down (fig.\,\ref{fig:Strahlengang}). Vacuum ultraviolet light was produced by a deuterium lamp with a MgF$_{\text{2}}$ window (Cathodeon Model V03). It was attached to an outer vacuum cell of 100\,mm diameter (CF 100 cross piece). An inner cell made from copper was installed in the center of this vacuum cell. This inner cell had two optical MgF$_{2}$ windows of 5\,mm thickness and 25\,mm diameter and could be filled with liquid argon.  The distance between the inner surfaces of the MgF$_{2}$ windows was 58\,mm, which is the relevant distance for the transmission measurements. The light emitted from the deuterium lamp was collimated by a stack of 3 apertures with 4\,mm diameter and sent through the center of the inner cell. Another set of apertures was mounted behind the inner cell. A f=30\,cm vacuum spectrometer (McPherson model 218) was mounted at the outer cell opposite to the deuterium lamp. It was equipped with a VUV image intensifier (micro channel plate, MCP) with MgF$_{2}$ entrance window, S20 multi-alkali cathode, and a diode array. The spectrometer could measure a spectrum of about 60\,nm width in one exposure, and the center wavelength could be set manually.
        
\begin{figure} 
            \includegraphics[width=\columnwidth]{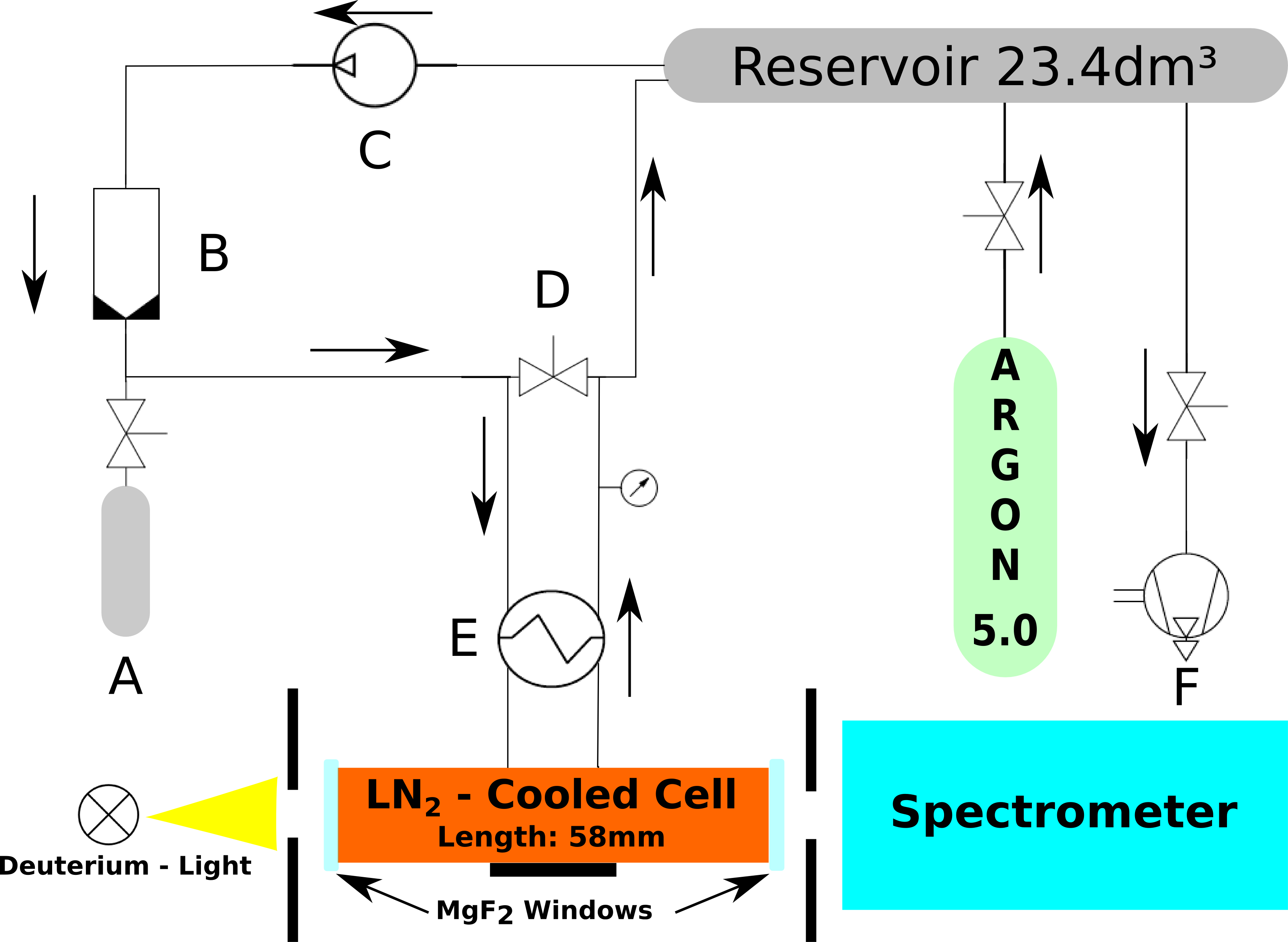} 
            \caption{\textit{Schematic drawing of the experimental setup. Broadband VUV light was sent from a deuterium lamp through the inner cell (see fig.\,\ref{fig:Strahlengang}), which could be either evacuated or filled with liquid argon, and detected by a spectrometer. The cell was operated as a bypass (valve D) of a closed gas system in which the argon gas was circulated with a metal bellows pump (C) through a rare gas purifier (B). A heat exchanger (E) was installed between the warm and cold parts of the gas system. A small storage volume (A) could be used to distill the gas prior to filling it into the system. A turbo molecular pump (F) was used to evacuate the system prior to filling.}}
            \label{fig:Prinzipskizze} 
        \end{figure}

        The cryogenic setup and gas handling were analogous to those described in refs.\,\cite{Heindl_1,Heindl_2}. The setup including a schematic drawing of the gas handling system is shown in fig.\,\ref{fig:Prinzipskizze}. The inner cell was held in place and cooled by a copper rod attached to a dewar filled with liquid nitrogen. Temperature was measured using PT100 resistors and could be controlled by sending an electrical current through a 47\,$\Omega$, 20\,W resistor, which was soldered onto the copper rod. Gas filling and circulation were performed via a heat exchanger. The outer gas system consisted of a closed cycle of a rare gas purifier (SAES model MonoToRR$^{\circledR}$ Phase II, PS4-MT3) and a metal bellows pump to circulate the gas through the system. The gas could be filled into the system from a gas bottle with a pressure reducer. The cryogenic cell was added as a bypass to the gas cycle so that the gas in that cell could be continuously refreshed by condensation and evaporation. A stainless steel storage volume was attached to the outer gas system, and could be cooled to liquid nitrogen temperature by immersing it into an open dewar filled with liquid nitrogen. This storage volume was used to distill the argon gas used for the absorption measurements as will be described below. Separate turbo molecular pumps were used to pump the gas system, the monochromator as well as the outer vacuum cell.
        
        \subsection{\label{subsec:Experimental_method}Experimental method}

        The basic concept was to divide the spectra obtained with the inner cell filled with liquid argon by the corresponding reference spectra recorded with the inner cell cooled and evacuated. Since both spectra were recorded with the inner cell at the same temperature, changes in the transmission due to geometrical effects (expansion and contraction of the rod holding the cell etc.) could be excluded. A fixed set of center wavelengths of 90, 120, 150, 180, and 210\,nm was used for the experiments described below. The individual spectra overlapped by 30\,nm. An overall spectrum was generated by minimizing the squared deviations of the optical signal in the overlapping regions \cite{Diplomarbeit}. This allowed us to measure absorption spectra from the short wavelength cutoff of the detection system up to 250\,nm above which the higher diffraction-order signals from the grating monochromator would have required edge filters to be used. The operation and readout of the detector was computer controlled and allowed to select the exposure time. Exposures could be repeated automatically several times to improve the statistics. The concept was to record transmission spectra when the inner cell was filled with liquid argon for every setting of the central wavelength and corresponding reference spectra with a cooled and evacuated inner cell. A necessary correction is to take the change in the index of refraction at the inner surfaces of the MgF$_{2}$ windows into account between the case of an empty cell and a cell filled with liquid argon (Fresnel formula). A test to correct our raw data with this Fresnel formula using the index of refraction of MgF$_{2}$ \cite{MgF2_Refractive_Index} and liquid argon (scaled from the gas phase) \cite{GAr_Refractive_Index,LAr_Refractive_Index}, however, did not change the overall spectral shape, in particular, the decreasing transmission towards shorter wavelengths (described in section \ref{subsec:Pure_liquid_argon}) remained. Therefore, the "Fresnel correction" was not applied.

\begin{figure} 
            \includegraphics[width=\columnwidth]{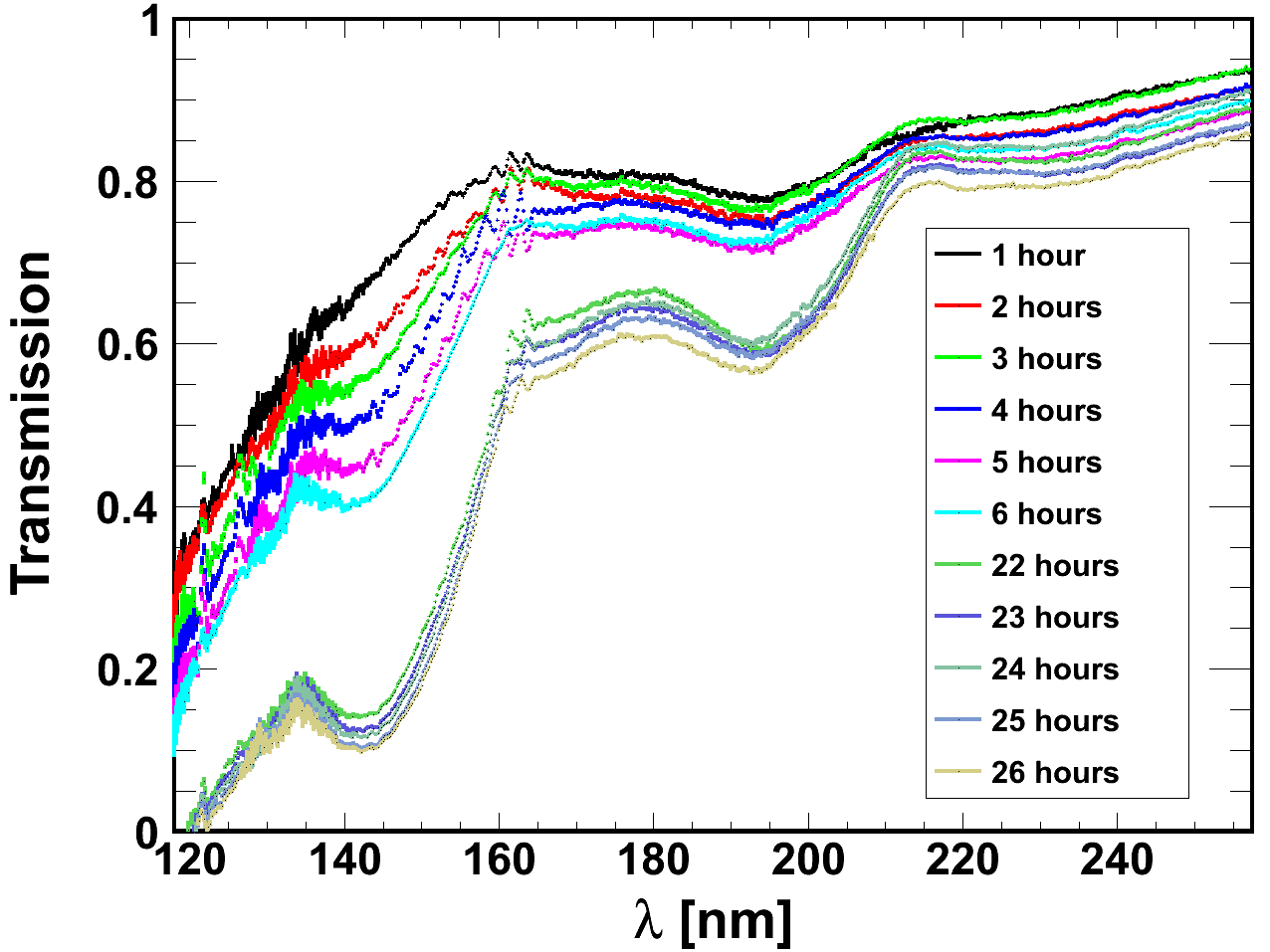} 
            \caption{\textit{The transmission of the empty and cooled inner cell is reduced with time by condensation of substances from the insulating vacuum of the outer cell onto the MgF$_{2}$ windows. A series of transmission spectra with respect to the time of filling the dewar with liquid nitrogen is shown. Spectra in the time interval from 2 to 3 hours were used to correct the data recorded for the actual transmission measurements of liquid argon. Spectra obtained with argon were typically measured 2 hours and reference spectra typically 3 hours after the start. Detailed time keeping in combination with interpolation (see fig.\,\ref{fig:Beschlag_expo_fit}) between the data shown above was performed to use the appropriate corrections to the data before calculating the transmission of liquid argon itself.}}
            \label{fig:Beschlag} 
        \end{figure}

\begin{figure}  
            \includegraphics[width=\columnwidth]{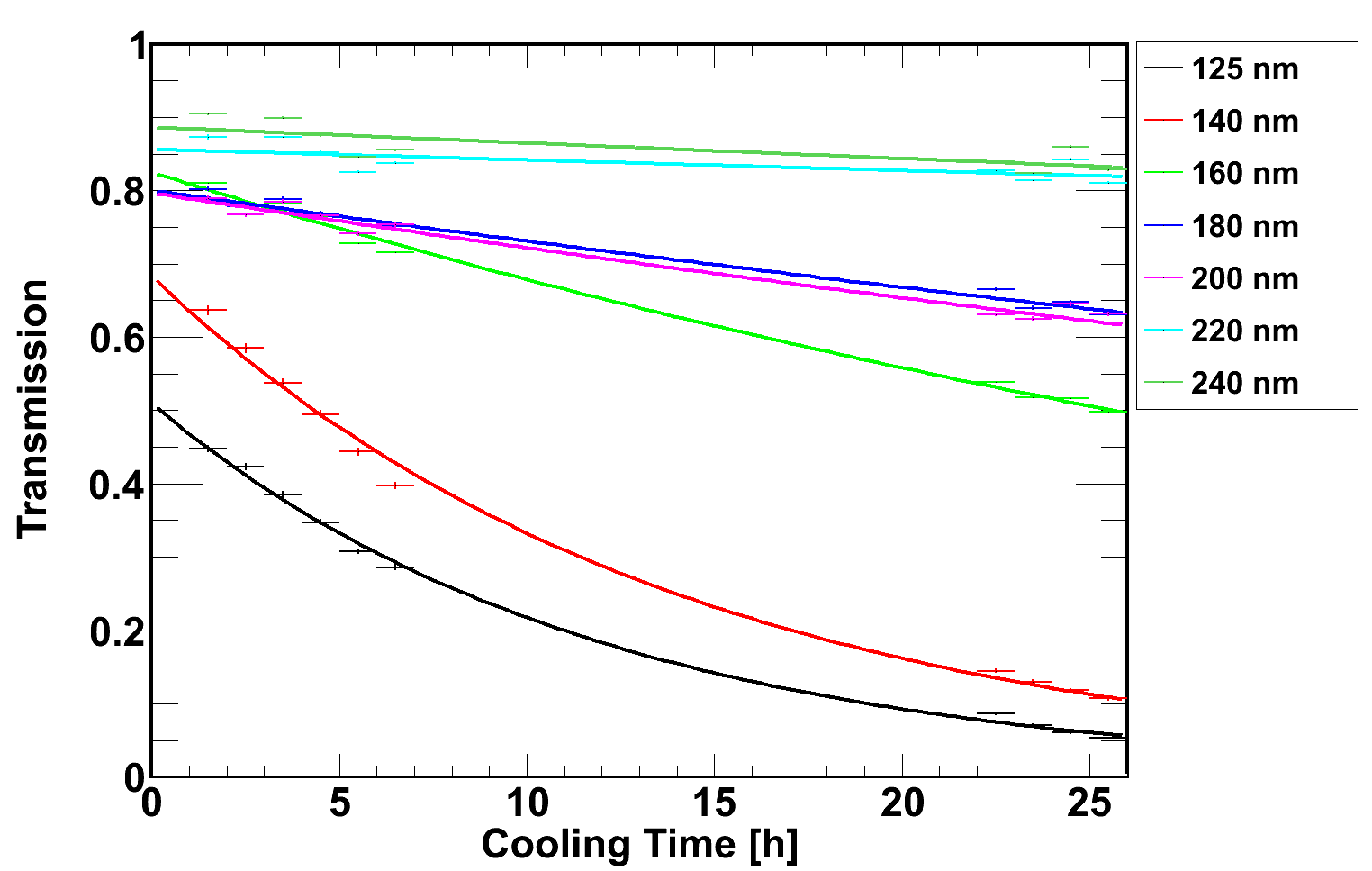} 
            \caption{\textit{Time dependence of the transmission due to "fogging" for a set of different wavelengths, together with exponential fits to the data.}}
            \label{fig:Beschlag_expo_fit} 
        \end{figure}

        It was found that a more serious experimental problem appears due to condensation of residual gas components in the outer vacuum onto the surfaces of the MgF$_{2}$ windows of the inner cell, a process which will be called "fogging" in the text below. A series of transmission spectra of the cooled and evacuated inner cell recorded at various times after cooling the inner cell down is shown in fig.\,\ref{fig:Beschlag}. An analysis of these data showed that the transmission was reduced exponentially with time at each wavelength position. This is documented in fig.\,\ref{fig:Beschlag_expo_fit} and allowed a proper correction of each spectrum according to the time at which it had been recorded after the start of the cooldown. The measurement of the time dependent fogging of the MgF$_{2}$ windows (fig.\,\ref{fig:Beschlag}) was performed 21 hours after the measurement of pure liquid argon. This was approximately the time difference which was necessary to heat up the cell to room temperature again, without thermally damaging the experimental setup. Therefore, changes in the composition of the residual gas components in the outer insulation vacuum between the measurement of pure liquid argon and the measurement of the fogging of the MgF$_{2}$ windows can be excluded.
        
        Since it was found that the liquid argon could be removed from the inner cell more quickly than condensed into it, the reference spectra were recorded after the spectra had been measured with the inner cell filled with the liquid. The time at which each spectrum was recorded with respect to the start of cooling the inner cell was carefully documented for corrections in the analysis of the data based on the findings shown in fig.\,\ref{fig:Beschlag} and fig.\,\ref{fig:Beschlag_expo_fit}. The accuracy of time measurement was of the order of seconds and the time differences between the measurements of the cell filled with liquid argon and the corresponding reference measurements were between one and two hours depending on the intensity of the D$_{2}$ lamp in the different spectral regions and consequently the required exposure times. The overall accuracy was of the order of minutes given by the exposure time of the data recording system. The statistical errors of the exponential fits of the transmission at each wavelength (fig.\,\ref{fig:Beschlag_expo_fit}) were considered and incorporated in the total error analysis. 

        Besides the short-term degradation of the spectra shown in figs.\,\ref{fig:Beschlag} and \ref{fig:Beschlag_expo_fit} a long term modification (weeks) of the emission from the deuterium lamp was found. Exposure to the residual gas in the outer cell obviously led to a reduction of intensity in the short wavelength part of the VUV emission due to carbon hydroxides in the residual gas, which were cracked by the VUV light and coated the MgF$_{2}$ window of the lamp. This effect is known in VUV optics and could be reversed by letting the lamp operate in an oxygen containing atmosphere for several hours. 

\begin{figure} 
            \includegraphics[width=\columnwidth]{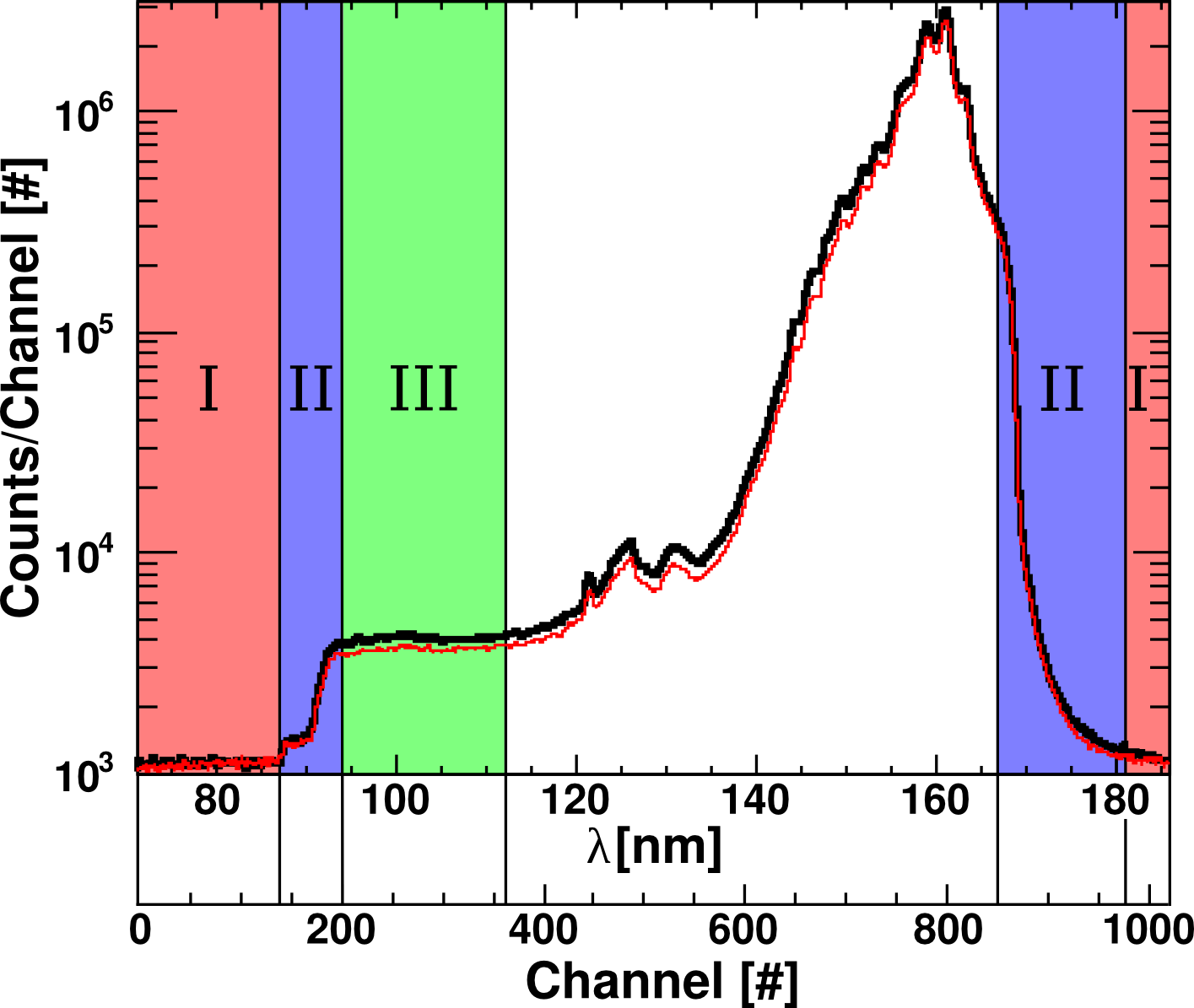} 
            \caption{\textit{An example of the raw data used in the experiments is shown. The thin (red) line shows the data from a measurement of the inner cell evacuated and the black line shows raw data from a measurement of the inner cell filled with liquid argon. The edge of the detector is not sensitive to light because the diode array is blocked by the housing of the intensifier (I, outer shaded area, red in colour online). After a transition phase (II, next shaded area, blue in colour online) the detector can be used for data taking between channels 200 and 850. Since no light is transmitted through the MgF$_{2}$ windows shorter than 113\,nm \cite{Abschnittkante_MgF2} the intensity level in the next shaded region on the left side (III, green in colour online) was used as the background for the data. In summary, it shows that the setup could be used above 118\,nm where the first real signal from the deuterium lamp can be observed. Note that in these raw data the reference spectrum (thin, red line) is below the signal spectrum (thin, black line) because neither "fogging" nor Fresnel corrections have been applied.}}
            \label{fig:Raw_data_Spectrum} 
        \end{figure}

        For the analysis of the data described below it was important to study the performance of the detector carefully. An important issue was to find the zero level in the spectra, which has to be subtracted from the raw data before the ratio between transmission spectrum and reference spectrum is calculated. First, background spectra were recorded by closing a valve in front of the spectrometer and subtracted from the subsequent measurements. Fig.\,\ref{fig:Raw_data_Spectrum} exhibits an example of the raw data measured in the most critical, short wavelength region (120\,nm central wavelength setting at the monochromator). The thin (red) line shows the data obtained in a measurement of the inner cell evacuated and the thick (black) line shows data recorded in a measurement with the inner cell filled with liquid argon. Relevant are two background levels in which scattered light is taken into account. The first pedestal (I in fig.\,\ref{fig:Raw_data_Spectrum}) which appears left and right in the spectra is due to the fact that the diode array is somewhat larger than the diameter of the MCP and sees no signal. After a transition phase (II in fig.\,\ref{fig:Raw_data_Spectrum}), the second step is observed in the region (III in fig.\,\ref{fig:Raw_data_Spectrum}) where the MCP and the diode array are both active. This level is used as a background level based on data which were recorded with wavelength settings for which no light can be expected: In our analysis we used wavelengths, which were shorter than the cutoff wavelength of the MgF$_{2}$ windows ($\lambda<$113\,nm \cite{Abschnittkante_MgF2}). A critical examination of the emission spectrum based on the emission spectrum of the lamp and the background issues of the detector showed that transmission measurements down to 118\,nm can be performed with the setup described here. The light output from the deuterium lamp below 118\,nm is too low to provide reliable data (see fig.\,\ref{fig:Raw_data_Spectrum}). Transmission spectra shown below will therefore always start at this wavelength.
        
\section{Results and discussion}
        
        \subsection{Pure liquid argon}
        \label{subsec:Pure_liquid_argon}
\begin{figure}
                    \includegraphics[width=\columnwidth]{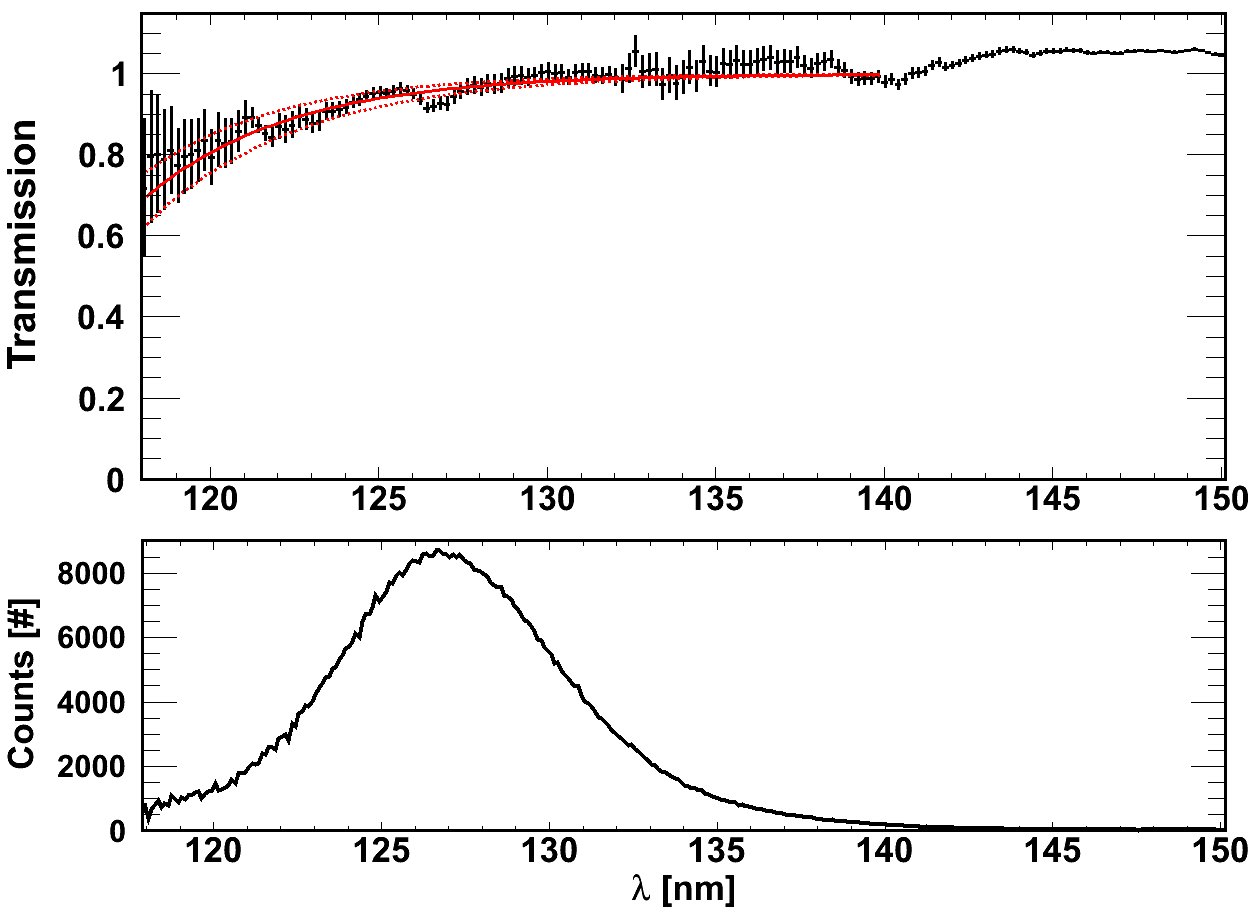} 
                    \caption{\textit{The transmission of well purified and distilled liquid argon (upper panel) is shown in comparison with the emission of liquid argon (lower panel) as taken from ref.\,\cite{Heindl_1}. The red line (upper panel) shows the exponential fit according to eq.\,\ref{eq:Expo_Fit}. The dashed red lines correspond to the two extreme combinations of the 1$\sigma$ errors of the fit parameters a and b (see text).}}
                    \label{fig:Transmission_Emission_LAr_purified_distillated} 
                \end{figure}

        \noindent
                The best result, obtained so far, for the wavelength dependence of the transmission of VUV light through 58\,mm liquid argon is shown in fig.\,\ref{fig:Transmission_Emission_LAr_purified_distillated} in comparison with the emission spectrum observed in ref.\,\cite{Heindl_1}. All purification steps described in section \ref{subsec:Effect_of_xenon_impurities} below (usage of a gas purifier and distillation of liquid argon) were applied to record these data. Since the raw data showed a transmission slightly above 100\% for long wavelengths, we compensated this so far unresolved systematic effect by scaling the 210\,nm center wavelength transmission measurements to 100\%. The transmission is close to 100\% above 130\,nm. It decreases towards shorter wavelengths and reaches 70\% at 118\,nm, the shortest wavelength of our measurements. It is believed that this decrease in transmission is due to argon itself and could be attributed to absorption by transitions corresponding to the analogue of the first argon excimer continuum. However, a residual unknown impurity, which in the presence of argon atoms causes this decrease, cannot be completely excluded. For an empirical description we fitted an exponential function (eq.\,\ref{eq:Expo_Fit}) with two parameters a and b to the data from $\lambda=118$\,nm to 140\,nm. The parameter b is a hint to the wavelength position of strong resonant absorption by the analogue of the first argon excimer continuum.
        
                \begin{equation}
                                    \frac{I(d,\lambda)}{I_{0}(\lambda)}=T(\lambda)=1-e^{-a\cdot(\lambda-b)} \label{eq:Expo_Fit} \\
                \end{equation}
                \begin{description}
                        \item [$I_{0}(\lambda)$] Incident light intensity as indicated by the spectrum (thin red line) in fig.\,\ref{fig:Raw_data_Spectrum}
                        \item [$I(d,\lambda)$] Light intensity after traversing a layer of thickness d
                \end{description}
                For our case, d=5.8\,cm, we obtain the following values for a and b together with their 1$\sigma$ errors: 
                \begin{eqnarray}
                                 a &=& \left(0.234\pm0.014\right)\,nm^{-1} \label{eq:Paramter_1}\\
                                 b &=& \left(113.02\pm0.65\right)\,nm \label{eq:Parameter_2}
                \end{eqnarray}

\begin{figure} 
                    \includegraphics[width=\columnwidth]{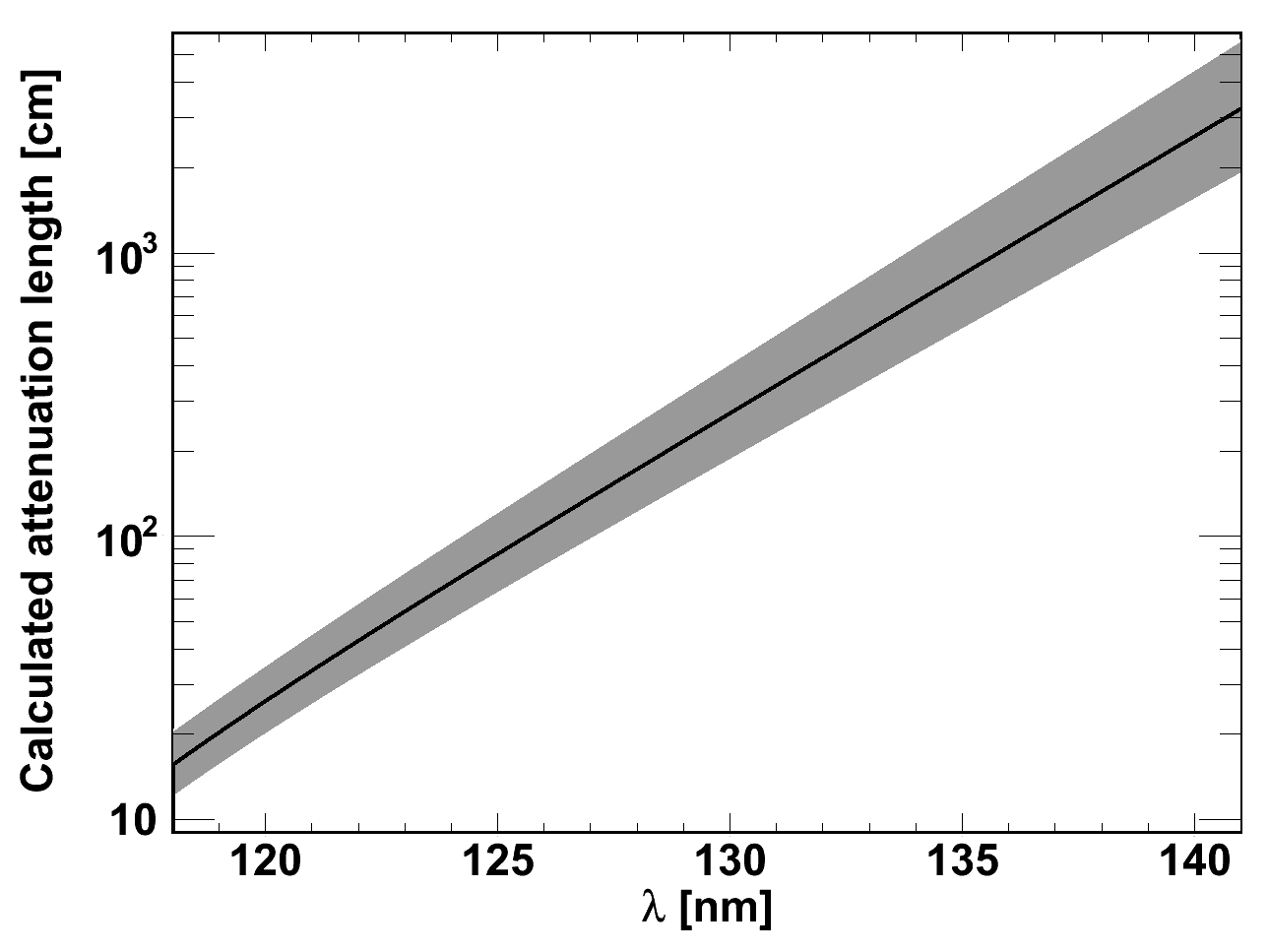} 
                    \caption{\textit{Wavelength-dependent attenuation length calculated from the fit in fig.\,\ref{fig:Transmission_Emission_LAr_purified_distillated}. The shaded area represents the error margins corresponding to the dashed lines in fig.\,\ref{fig:Transmission_Emission_LAr_purified_distillated}. Above 140\,nm the calculated attenuation length diverges due to the transmission approaching unity.}}
                    \label{fig:Attenuation_length_log} 
                \end{figure}

\begin{figure} 
                    \includegraphics[width=\columnwidth]{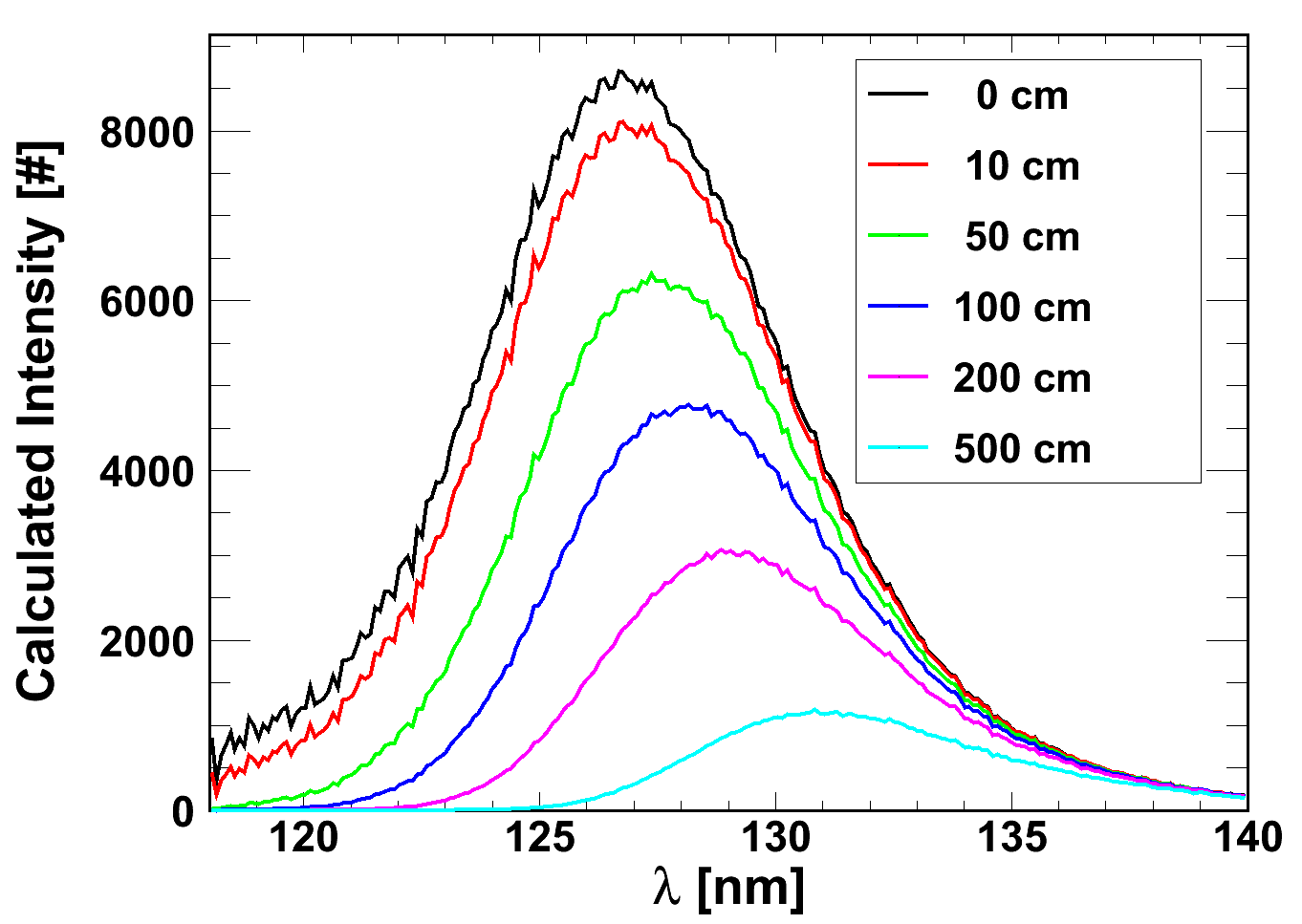} 
                    \caption{\textit{The modification of the fluorescence spectrum of liquid argon when the light has traversed the distances indicated in the insert  is shown. A significant "reddening" of the spectrum occurs due to the wavelength dependence of the transmission data in addition to the overall light attenuation.}}
                    \label{fig:Emission_different_lengths}  
                \end{figure}

                The transmission data and the fit (full red line) are shown in fig.\,\ref{fig:Transmission_Emission_LAr_purified_distillated} (upper panel). The dashed (red) lines correspond to the two extreme combinations of the 1$\sigma$ errors of the fit parameters a and b. The experimental method as described in section\,\ref{subsec:Experimental_method} leads to transmission values slightly above unity around $\lambda$=145~nm. This is caused by a systematic error, which cannot be resolved until a measurement of the wavelength dependent refractive index of liquid argon is available to provide a correct reflection value at the surface between liquid argon and the MgF$_{2}$ windows (Fresnel correction). Consequently, the presented transmission curve should rather be seen as an upper limit for the real transmission. However, the decrease towards the short wavelength end of the presented transmission spectrum (see fig.\,\ref{fig:Transmission_Emission_LAr_purified_distillated}) is visible in all measurements and cannot be attributed to a systematic error. This decrease is probably even underestimated, due to the optimistic transmission scaling described above. The empirical description for the wavelength dependence of the transmission (eq.\,\ref{eq:Expo_Fit}) is now used to calculate the wavelength dependent attenuation lengths ($\lambda_{att}(\lambda)$). The attenuation of light, traversing a medium can be described by an exponential decay\footnote{Equation\,\ref{eq:Light_attenuation} neglects any potential flourescence-light emission. This is expected to be justified since the emission of the deuterium lamp is weak below about 130~nm and the distance between the inner cell and the detector is large ($\sim45$~cm).} at each wavelength $\lambda$:
                \begin{eqnarray}
                                                                                           I(d,\lambda) &=& I_{0}(\lambda) \cdot e^{-\frac{d}{\lambda_{att}(\lambda)}} \label{eq:Light_attenuation} \\
                                        \xrightarrow{\ \ \ } \lambda_{att}(\lambda) &=& \frac{d}{\text{ln}\left(\frac{I_{0}(\lambda)}{I(d,\lambda)}\right)} \label{eq:Attenuation_length} \\
                        \xrightarrow{eq.\,\ref{eq:Expo_Fit}} \lambda_{att}(\lambda) &=& \frac{d}{\text{ln}\left(\frac{1}{T(\lambda)}\right)} \label{eq:Attenuation_length_Fitfunction}
                \end{eqnarray}
                Using eq.\,\ref{eq:Attenuation_length_Fitfunction} the wavelength dependent attenuation length was calculated. The result is shown in fig.\,\ref{fig:Attenuation_length_log}.
                Note, that no unique attenuation length can be given for liquid argon in the VUV spectral range. However, combining a source spectrum with the transmission data, the wavelength integrated intensity versus distance traversed by the light in liquid argon can be calculated. This is demonstrated using the important example when fluorescence light of argon itself (fig.\,\ref{fig:Transmission_Emission_LAr_purified_distillated} lower panel) is attenuated and thereby modified in its spectral shape during its path through liquid argon. The modification of the spectral shape with distance is calculated using eq.\,\ref{eq:Light_attenuation} with the obtained attenuation length $\lambda_{att}(\lambda)$ from eq.\ \ref{eq:Attenuation_length_Fitfunction} and is depicted in fig.\,\ref{fig:Emission_different_lengths}. It shows the "reddening" of VUV light propagating in liquid argon just like sunlight observed around sunset. Wavelength integrated data for the light flux F(x) versus distance can be calculated from spectra analogous to fig.\,\ref{fig:Emission_different_lengths} along the path of light using eqs.\,\ref{eq:Integral_Calculation_1} and \ref{eq:Integral_Calculation_2}, thereby taking the modification of the spectrum into account:
                
                \begin{eqnarray}
                   \frac{F(x)}{F_{0}}&=&\frac{\int\limits_{118\,nm}^{140\,nm}I(x,\lambda)\ d\lambda}{\int\limits_{118\,nm}^{140\,nm}I_{0}(\lambda)\ d\lambda}  \label{eq:Integral_Calculation_1}\\ 
                                         &=&\frac{\int\limits_{118\,nm}^{140\,nm}I_{0}(\lambda) \cdot e^{-\frac{x}{\lambda_{att}(\lambda)}}\ d\lambda}{\int\limits_{118\,nm}^{140\,nm}I_{0}(\lambda)\ d\lambda} \label{eq:Integral_Calculation_2}
                \end{eqnarray}
                
\begin{figure} 
                    \includegraphics[width=\columnwidth]{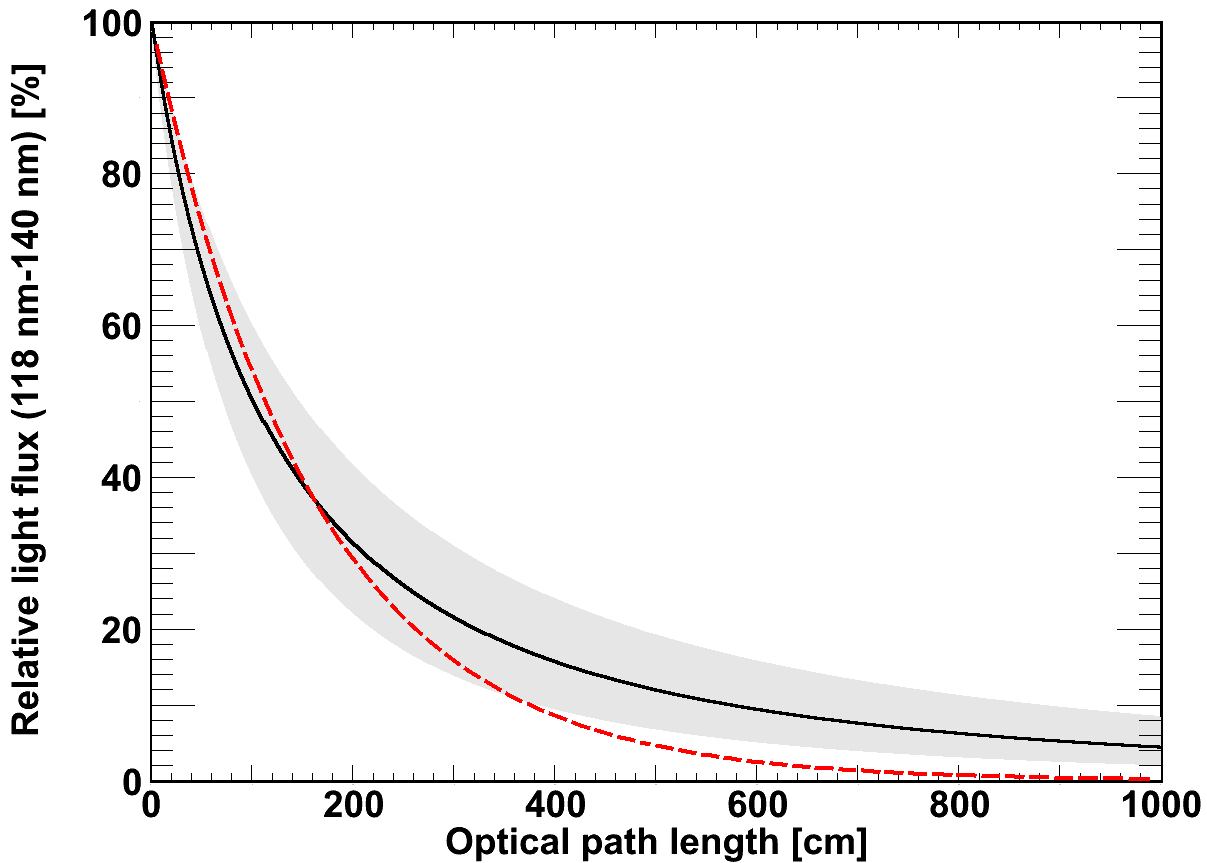} 
                    \caption{\textit{The attenuation of wavelength-integrated argon fluorescence light (light flux) with distance traversed in liquid argon (black line, see eq.\,\ref{eq:Integral_Calculation_2}) is shown in comparison with an exponential attenuation of light (dashed red line). Although the flux reduction due to the wavelength dependent absorption coefficients is not exponential, the $\frac{1}{e}$ attenuation-length value of 1.63~m is indicated as point of intersection of the two lines for comparison with experiments where a wavelength independent attenuation is assumed. The shaded area indicates the error margins corresponding to fig.\,\ref{fig:Attenuation_length_log}.}}
                    \label{fig:Integral_Emission_118_140} 
                \end{figure}

                The results are shown in fig.\,\ref{fig:Integral_Emission_118_140}. As mentioned above no unique attenuation length exists in liquid argon. The flux decrease with distance depends on the combination of the spectrum at the starting point and its modification by the wavelength dependent attenuation, and has to be calculated for each wavelength independently. Wavelength integrated values have to be obtained by integrating the spectrum afterwards and comparing this integral with the integral of the spectrum at the starting point. The flux reduction is not exponential as the comparison with an exponential function in fig. \ref{fig:Integral_Emission_118_140} shows. The nominal $\frac{1}{\text{e}}$ value with respect to the initial flux of the argon fluorescence light is 1.63\,m.
                
        \subsection{\label{subsec:Effect_of_xenon_impurities}Effect of xenon impurities}
\begin{figure} 
                    \includegraphics[width=\columnwidth]{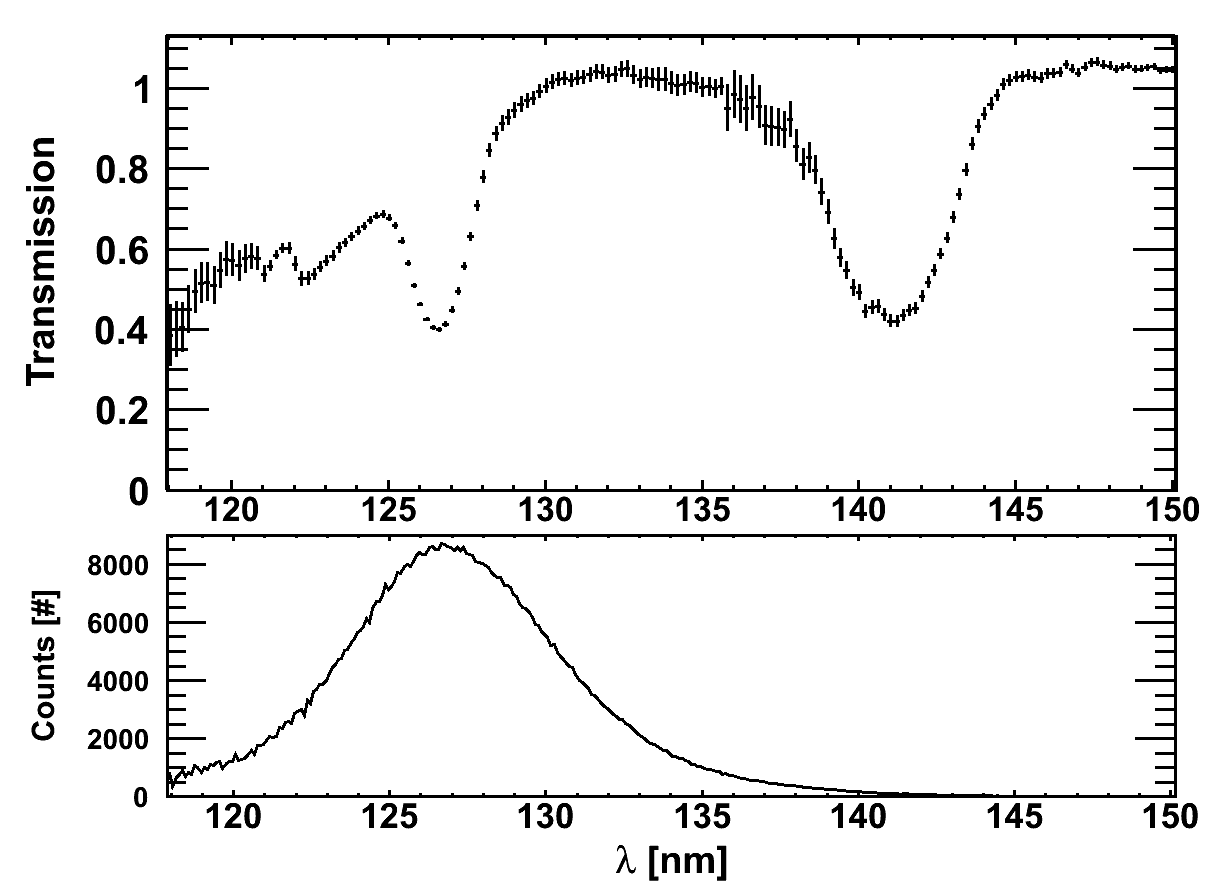} 
                    \caption{\textit{The transmission of chemically purified but not distilled liquid argon (upper panel) is shown in comparison with the fluorescence spectrum of liquid argon (lower panel \cite{Heindl_1}). The two pronounced absorption dips at 126.5 and 141.0\,nm are related to xenon impurities, as described in detail in ref.\,\cite{Jortner_1970}. Note that the xenon absorption feature at 126.5\,nm is in almost exact coincidence with the peak of the argon emission spectrum. }}
                    \label{fig:Transmission_Emission_LAr_purified_undistillated} 
                \end{figure}

        \noindent
                Although the 58\,mm length of the absorption cell was too short for measuring reliable transmission values for wavelengths longer than about 130\,nm it had the right length for studying the effect of impurities and gas handling. The transmission spectra shown in figs.\,\ref{fig:Transmission_Emission_LAr_purified_distillated} and \ref{fig:Transmission_Emission_LAr_purified_undistillated} could only be observed after a series of about 20 measuring campaigns with stepwise improved gas handling. Here we follow this way backwards to emphasize the most critical issues by describing them first. Already in the emission spectra of ref.\,\cite{Heindl_1} it was found that it is difficult to remove other rare gases from argon, xenon in particular. Why krypton never played a role in our experiments cannot be answered by the authors, so far. In general, xenon (or may be in some cases krypton) may be contained in some gas bottles, or it may be released from components in the gas system if it had been used with xenon or krypton prior to argon measurements. Poly-tetra-fluor-ethylene (PTFE, Teflon) is one of the materials, which is critical because it is often used for gaskets, e.g. in the connector of the pressure reducer to the gas bottle. Measurements exposing a 1\,mm thick Teflon sheet of 16\,g weight for 1\,h to 2\,bar xenon led to an increase of 25\,mg in weight, which was released again with a time constant of 2 to 3 days when it was again exposed to ambient air. This behaviour is also in agreement with the observation that another rare gas pressure reducer, used on a xenon bottle more than a year ago, still led to xenon impurities of more than 50\,ppm even after several purgings. 
                
\begin{figure} 
                    \includegraphics[width=\columnwidth]{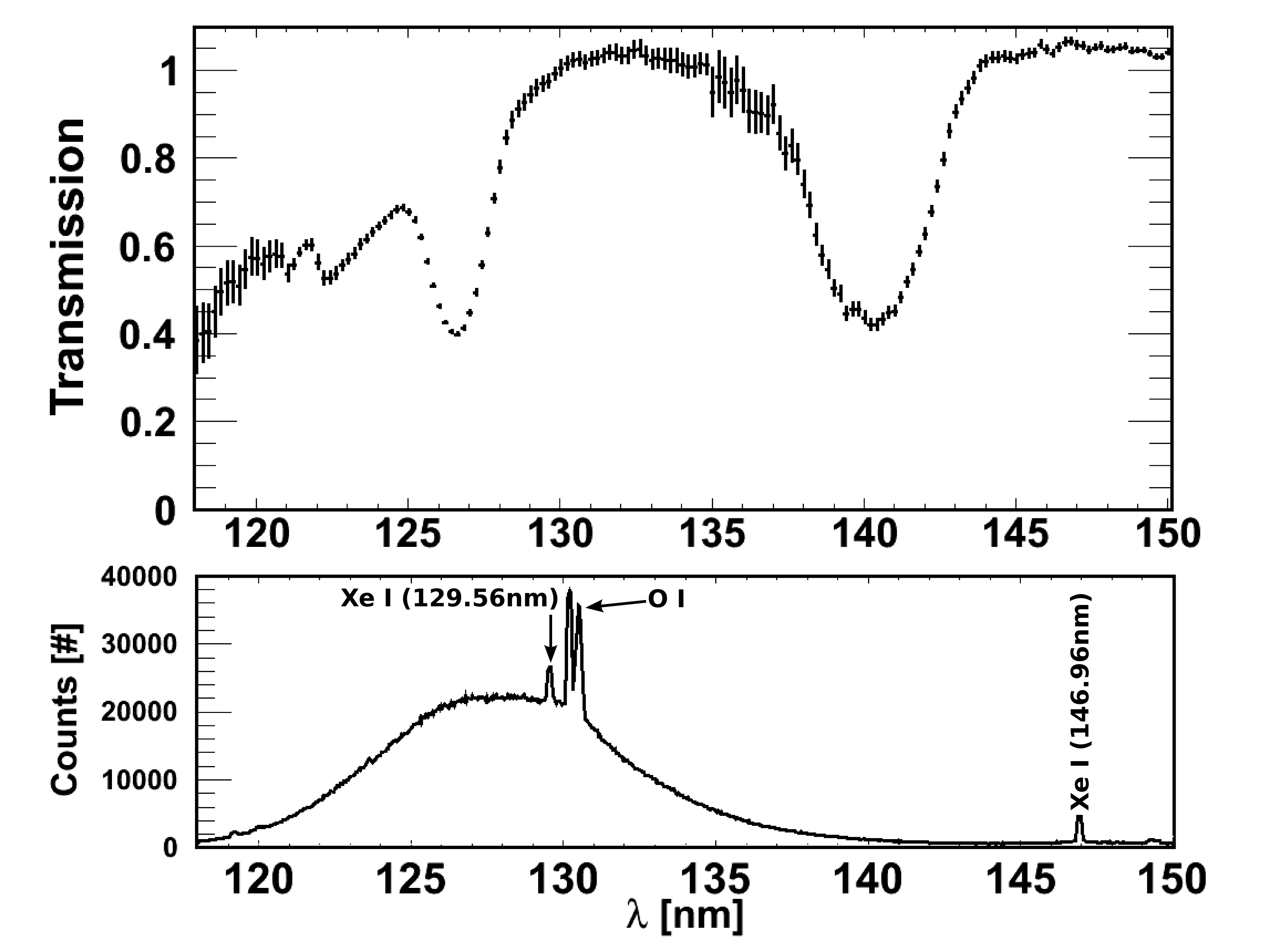} 
                    \caption{\textit{Comparison of the transmission of chemically purified but not distilled liquid argon (upper panel) and the emission spectrum of unpurified gaseous argon taken directly from the gas bottle (lower panel), which was used in our experiments, at 900\,mbar and a temperature of 20\textdegree C. The lines at 130.2 and 130.5\,nm can be attributed to an oxygen contamination (neutral oxygen O\,I). The other emission lines are the resonance lines of neutral xenon (Xe\,I) at 129.56 and 146.96\,nm.}}
                    \label{fig:Transmission_Emission_ArXe} 
                \end{figure}

                The problem is that xenon (or krypton if it were contained in a gas system) cannot be removed by regular rare gas purifiers because they make use of chemical reactions and can therefore not act on rare gases. The situation is even more critical in experiments with a cryogenic cell because xenon (or krypton) impurities will condense in the experimental cell with time from all parts of the gas system at higher temperatures. Absorption features which could be attributed to xenon were indeed observed in the experiments described here. In fig.\,\ref{fig:Transmission_Emission_LAr_purified_undistillated} (upper panel) the transmission of a liquid argon sample is shown for which the conventional chemical purification had been applied but the gas was not distilled before being used for the measurement. The fluorescence of liquid argon is again shown for comparison (lower panel). This demonstrates that there is, unfortunately, an almost exact coincidence between one of the xenon absorption features at 126.5\,nm and the peak of the argon emission spectrum. For the assignment of the absorption dips a comparison of the absorption spectrum with the emission spectrum of our unpurified argon gas is shown in fig.\,\ref{fig:Transmission_Emission_ArXe}. A low-energy electron beam excitation method described in ref.\,\cite{E_beam_induced_light_emission} and references therein was used to record the emission spectrum in the gas phase. The resonance lines of xenon at 146.96 and 129.56\,nm \cite{NIST} are clearly visible (see fig:\,\ref{fig:Transmission_Emission_ArXe}, lower panel). According to ref. \cite{Jortner_1970} the pronounced absorption dip at 141.0\,nm can be attributed to a perturbed atomic transition of xenon ($^{1}\text{S}_{0}$ $\rightarrow$ $^{3}\text{P}_{1}$, 146.96\,nm). The second absorption dip at 126.5\,nm, however, cannot be assigned to a perturbed atomic transition but rather to a trapped exciton (Wannier-Mott type) impurity state \cite{Jortner_1970}. The widths of the absorption bands are 2.0 and 3.8\,nm, respectively. Xenon is therefore a very critical impurity, which can as a pure rare gas be easily underestimated in practical applications. 
                
                The concentration of xenon in the argon gas used in the present experiments could be estimated from emission measurements performed in the gas phase (fig.\,\ref{fig:Transmission_Emission_ArXe} lower panel) to be smaller than 3\,ppm. This is obtained from a comparison with data shown in ref.\,\cite{Efthimiopoulos}. The concentration of xenon could also be estimated from a comparison with measurements by Raz and Jortner \cite{Jortner_1970}. There, the optical density of liquid argon doped with 3\,ppm of xenon was measured in a wavelength range from 120\,nm to 160\,nm, and for an optical path length of 1\,cm. The transmission minimum at 141.0\,nm is also clearly visible. A recalculation of their data to an optical path length of 5.8\,cm and a subsequent comparison of the areas of the transmission minima at 141.0\,nm lead to a concentration of approximately 140\,ppb xenon for the undistilled argon sample used in our experiment. The effect of the fractional distillation can also be estimated. A comparison of the area of the small transmission minimum in pure liquid argon at 141.0\,nm (fig.\,\ref{fig:Transmission_Emission_LAr_purified_distillated} upper panel), which is still visible but reduced compared to the undistilled liquid argon (e.g. fig. \ref{fig:Transmission_Emission_LAr_purified_undistillated} upper panel) leads to a concentration of approximately 7\,ppb. Consequently, the xenon contained in the argon was reduced by a factor of approximately 20 due to fractional distillation.
        
        \subsection{Absorption in unpurified liquid argon}

\begin{figure} 
                    \includegraphics[width=\columnwidth]{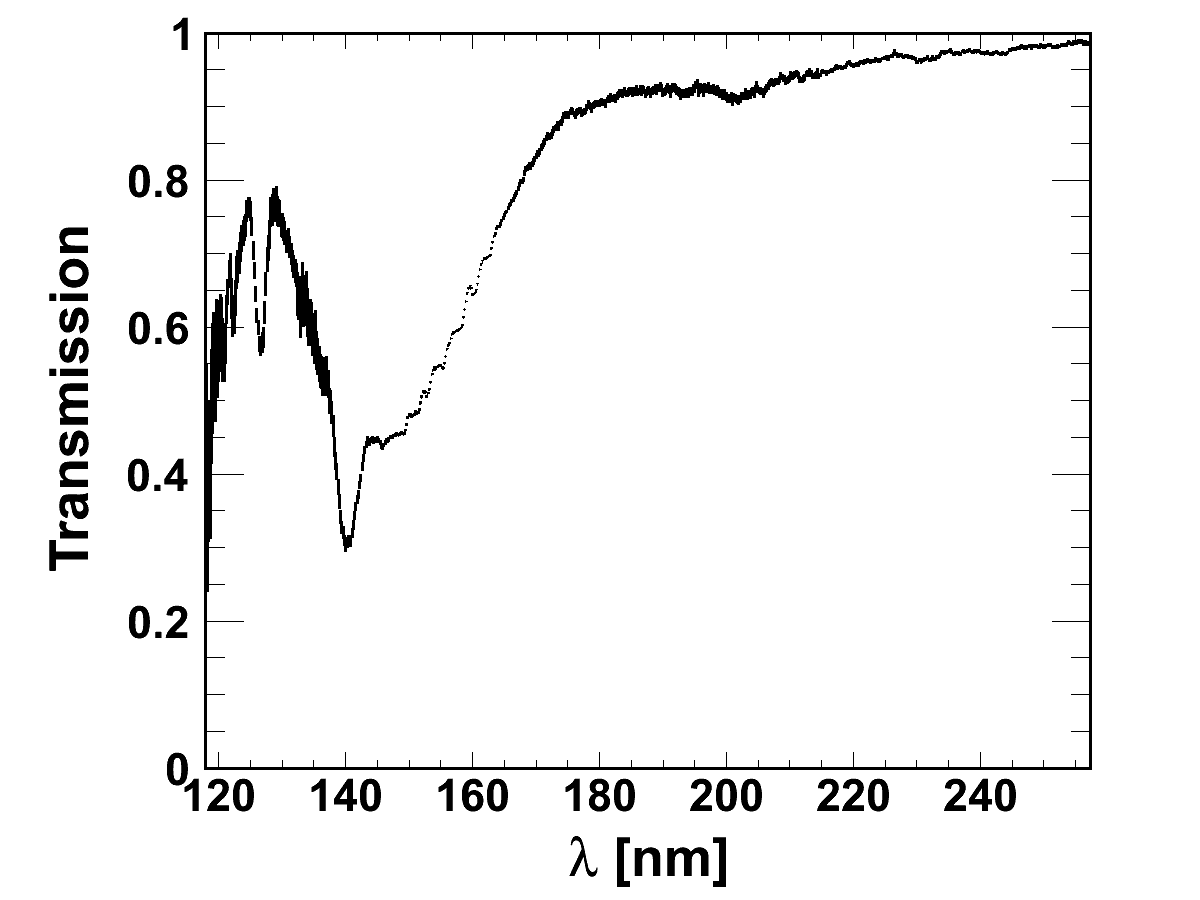} 
                    \caption{\textit{The transmission of argon with a nominal purity of 99.999\% is shown versus wavelength. A broad absorption feature can be observed below 180\,nm besides the two narrow xenon absorption bands shown in figs. \ref{fig:Transmission_Emission_LAr_purified_undistillated} and \ref{fig:Transmission_Emission_ArXe}. The broad structure is believed to be related mainly to water ice on the windows, coming from residual water in the argon.}}
                    \label{fig:Nicht_gereinigt_nicht_destilliert} 
                \end{figure}

                A transmission spectrum of liquid argon as taken directly from the gas bottle (Ar 5.0, i.e. a nominal purity of 99.999\,\%) is shown in fig. \ref{fig:Nicht_gereinigt_nicht_destilliert}. Besides the two absorption bands at 141.0 and 126.5\,nm, which have been attributed to a xenon impurity, a broad absorption is observed below 180\,nm with a minimum transmission of about 45\% at 145\,nm. The absorption/emission structure around 121\,nm may be due to hydrogen released from the water or an artifact related to the Lyman-$\alpha$ line of hydrogen emitted by the deuterium lamp. The broad absorption between about 125 and 180\,nm is believed to be mainly due to water vapor contained in the argon gas, which has not been circulated through the rare gas purifier. The temperature dependent absorption of solid water in the VUV had been studied in refs. \cite{Ice_crystall_formation,Absorption_Ice}. 
                
                It has to be noted that the transmission shown in fig.\,\ref{fig:Nicht_gereinigt_nicht_destilliert} was only reached when the system had already been used for a long time with many purification cycles with the rare gas purifier. In the beginning of the experiments the transmission reduced towards shorter wavelengths starting at 180\,nm reaching zero at about 150\,nm. This means that chemical gas purification is mandatory for fluorescence detectors using liquid argon as detector material. 
                
                Finally, it should be noted that the transmission measurement had in one test experiment been extended to the visible spectral range. The data are not easy to interpret because of the higher diffraction orders of the grating in the VUV monochromator. However, since all structures in the spectra appeared at wavelength positions which could be related to first order structures we can conclude that no significant absorption exists in liquid argon in the longer wavelength UV and visible spectral range.  

\section{Summary}

        It has been shown that the light propagation in liquid argon, which is frequently and in large quantities used for particle detection, is strongly wavelength dependent. However, a purification level can be reached where an effective attenuation length on the order of 1.6\,m can be obtained for its own fluorescence emission. If much longer propagation lengths are possible will become clear when experiments with higher accuracy will be performed, using longer absorption cells and improved light sources.

        We found a systematic reduction of transmission in liquid argon from about 140\,nm towards shorter wavelengths and attribute this to absorption in argon by transitions which are called the "first continuum" in emission. Note that the atomic and molecular structure of the heavier rare gases neon through xenon are so similar that essentially all effects measured for one of the species can also be found in the other ones. Therefore, an absorption feature analogous to the one observed here must also be expected, e.g., in particle detectors using liquid xenon as the scintillating material.

        Special care has always to be taken to avoid impurities. This is common knowledge as far as regular chemical rare gas purification is concerned. In the present paper it is shown that other rare gases, xenon in particular, lead to absorption bands, which in the case of xenon perfectly coincide with the emission band of argon. Therefore, it is important to test liquid argon used as scintillation material for the presence of this impurity. 

\section{Outlook}

        It may be worthwhile to explore the region near the wavelength cutoff of the present experiment (120 to 130\,nm) with better statistics. A light source described in ref.\,\cite{Dandl} emitting strongly in this region could be used for that purpose. Experiments with synchrotron radiation could also be considered \cite{Synchrotron}. Critical opalescence (in dense gases), Rayleigh scattering (in general), and fluctuations of the index of refraction driven by temperature gradients etc. (in the liquid phase) will have to be considered in experiments with longer optical path lengths than that used in the present experiment. 

        Xenon and krypton added deliberately in well defined quantities could provide information about the absorption features of mixed Ar-Xe and Ar-Kr molecules and their absorption spectra. Absorption and emission experiments in the liquid and the gas phase could be performed for such a study. As an experimental improvement cryogenic baffles in front of the inner cell and cooled, e.g., to liquid helium temperature should be installed to avoid the degradation of the transmission of the inner cell's optical windows by condensation of impurities from the insulating vacuum surrounding the inner cell.

        Finally, we are aware of the fact that a setup in which monochromatic light would be used instead of the broadband emission of a deuterium lamp would be superior to the setup presented here because possible problems with fluorescence induced by the probe light could then be avoided. The light intensity at the exit of a monochromator, however, would have been too weak and the light would have had a divergence related to the aperture ratio of the monochromator, which would have required a much more complex optical setup. The issue of fluorescence will be tested in a separate experiment by trying to observe light through an additional window of the inner cell through which scattered or reemitted light can be observed at 90\textdegree \ with respect to the incident light.
        
\section*{Acknowledgements}
We thank Roman Gernh\"auser for kindly providing the gas system.
This work was supported by funds of the Maier-Leibnitz-Laboratorium (Garching).


\begin{thebibliography}{99}
\bibliographystyle{unsrt}
  \bibitem{MiniCLEAN}
        A.\,Hime (MiniCLEAN Collaboration),\\ 
        arXiv:1110.1005v1 [physics.ins-det] (2011).

  \bibitem{DEAP/CLEAN}
        K.S.\,Olsen, 
        arXiv:0906.0348v1 [hep-ex] (2009)

  \bibitem{ArDM_1}
        A.\,Rubbia et al.,
        J. Phys.: Conf. Ser. \textbf{39}, 129 (2006)

  \bibitem{ArDM_2}
        A.\,Marchionni et al.,
        J. Phys.: Conf. Ser. \textbf{308}, 012006 (2011)

  \bibitem{WARP}
        P.\,Benetti et al.,
        arXiv:astro-ph/0701286v2 (2007)

  \bibitem{DarkSide}
        A.\,Wright (DarkSide Collaboration),\\
        arXiv:1109.2979v1 [physics.ins-det] (2011)

  \bibitem{DARWIN_1}
        L.\,Baudis (DARWIN consortium),\\
        arXiv:1012.4764v1 [astro-ph.IM] (2009)

  \bibitem{DARWIN_2}
        M.\,Schumann (DARWIN consortium),\\ 
        arXiv:1111.6251v1 [astro-ph.IM] (2011)

  \bibitem{ICARUS}
        C.\,Rubbia et al.,
        JINST \textbf{6} P07011 (2011)

  \bibitem{Rubbia_CP_Violation}
        A.\,Rubbia,
        arXiv:hep-ph/0402110v1 (2004)

  \bibitem{GERDA}
        A.A.\,Smolnikov (GERDA Collaboration),\\ 
        arXiv:0812.4194v1 [nucl-ex] (2008)

  \bibitem{Ishida_Abschwaechlaenge}
        N.\,Ishida et al.,
        Nucl. Instrum. Meth. A \textbf{384}, 380 (1997)

  \bibitem{Heindl_1}
        T.\,Heindl et al.,
        EPL \textbf {91}, 62002 (2010)

  \bibitem{Jortner_1970}
        B.\,Raz and J. Jortner,
        Proc. Roy. Soc. A \textbf{317}, 113 (1970)

  \bibitem{Photoabsorption_Cross_Section_Krypton_Xenon}
        J.B.\,Gerardo and A.\,W.\,Johnson,
        Phys. Rev. A \textbf{10}, 4 (1974)

  \bibitem{Rayleigh_scattering_length_calculation}
        G.M.\,Seidel, R.E. Lanou, and W. Yao,
        Nucl. Instrum. Meth. A \textbf{489}, 189 (2002)

  \bibitem{NIST}
         Y.\,Ralchenko et al. (NIST ASD Team),
         National Institute of Standards and Technology, Gaithersburg, MD (2011)

  \bibitem{Heindl_2}
        T.\,Heindl et al.,
        JINST \textbf {6}, P02011 (2011)
  
  \bibitem{Diplomarbeit}
        A.\,Neumeier,
        Diploma Thesis,\\ Technische Universit\"at M\"unchen (2012)

  \bibitem{MgF2_Refractive_Index}
        P.\,Laporte et al.,
        J. Opt. Soc. Am. \textbf {73}, 8 (1983)

  \bibitem{GAr_Refractive_Index}
        A.\,Bideau-Mehu et al.,\\
        J. Quant. Spectrosc. Radiat. Transfer \textbf {25}, 395 (1981)

  \bibitem{LAr_Refractive_Index}
        M.\,Antonello et al.,\\
        Nucl. Instrum. Meth. A \textbf {516}, 348 (2004)

  \bibitem{Abschnittkante_MgF2}
        W.\,R.\,Hunter, "Windows and Filters" in J.\,A.\,R. Samson and D.\,L.\, Ederer (Ed.) "Vacuum Ultraviolet Spectroscopy I", Academic Press San Diego, London, Boston, New~York, Sydney, Toronto (1998)

  \bibitem{E_beam_induced_light_emission}
        A.\,Ulrich et al.,
        Eur. Phys. J. Appl. Phys. \textbf {47}, 22815 (2009)

  \bibitem{Efthimiopoulos}
        T.\,Efthimiopoulos, D. Zouridis, and A. Ulrich,\\
        J. Phys. D: Appl. Phys. \textbf{30}, 1746 (1997)

  \bibitem{Ice_crystall_formation}
        M.\,Blackman and N.\,D.\,Lisgarten,\\
        Proc. Roy. Soc. \textbf{239}, 93 (1957)

  \bibitem{Absorption_Ice}
        R.\,Onaka and T. Takahashi,\\
        J. Phys. Soc. Jap. \textbf {24}, 548 (1968)

  \bibitem{Dandl}
        T.\,Dandl et al.,
        EPL \textbf{94}, 53001 (2011)

  \bibitem{Synchrotron}
        T.\,M\"oller et al.,
        Chem. Phys. Lett. \textbf{117}, 301 (1985)
\end{thebibliography}
\end{document}